\documentstyle{elsart} 
\begin{document} 
\begin{frontmatter}
\title{Gravitational Analog of the Electromagnetic Poynting Vector}

\author{L.M. de Menezes \thanksref{email}}
\thanks[email]{E-mail:demenezes@uvphys.phys.uvic.ca} \address{Dept. of
Physics and Astronomy, University of Victoria, Victoria, B.C. Canada
V8W 3P6}

\begin{abstract}
The gravitational analog of the electromagnetic Poynting vector is
constructed using the field equations of general relativity in the
Hilbert gauge. It is found that when the gravitational Poynting vector
is applied to the solution of the linear mass quadrupole oscillator,
the correct gravitational quadrupole radiation flux is
obtained. Further to this, the Maxwell-like gravitational Poynting
vector gives rise to Einstein's quadrupole radiation formula. The
gravitational energy-momentum (pseudo) tensor obtained is symmetric
and traceless. The former property allows the definition of angular
momentum for the free gravitational field.
\end{abstract}
\end{frontmatter}

\section{Introduction}
Despite the rather different frame works from which they arise, the
equations of general relativity have many similarities with those of
the electromagnetic field. These similarities are seen in the weak
field limit of general relativity where the geodesic equations of
motion and the field equations have essentially the same structure as
the equations of electromagnetic theory. In this work, general
relativity and electromagnetism are brought one step closer. It is
shown that in the Hilbert gauge, the field equations of general
relativity can be generated from a Lagrangian density similar in
structure to that of electrodynamics. The structure of the
energy-momentum (pseudo) tensor obtained from this Lagrangian density
coincides with that of the free electromagnetic field: it is
symmetric, traceless and can be constructed from the electromagnetic
energy-momentum tensor by replacing the Maxwell tensor $F^{\mu\nu}$ by
a suitable antisymmetric gravito-electromagnetic (pseudo) tensor
$\Pi^{\mu\nu}$. It is remarkable that the simple Maxwell-like
gravitational (pseudo) tensor gives rise to Einstein's quadrupole
radiation formula. To the author's knowledge this work represents a
novel approach to the study of gravitational fields in the weak field
limit.

\section{Electromagnetism and gravity}
 The electromagnetic action \cite{Landau}:
\begin{equation}
S\;=\;-\sum{\int}{m}\d s-\int{A_{\mu}J^{\mu}}\d\Omega -
\frac{1}{16\pi}\int{F_{\mu\nu}F^{\mu\nu}}\d\Omega 
\end{equation}
gives rise to the first and second pair of Maxwell's equations
\begin{equation}
F_{\mu\nu,\alpha}+F_{\alpha\mu,\nu}+F_{\nu\alpha,\mu} = 0,\;\;F^{\mu\nu}_{,\nu} = -\;4\pi{J^{\mu}},    \label{mfe}
\end{equation}
where
\begin{equation}
F_{\mu\nu}\;=\;A_{\nu,\mu}-A_{\mu,\nu},\;\;
J^{\mu}\;=\;(\rho,\rho\;{\bf v}). \label{mtensor}  
\end{equation}
In $(\ref{mtensor})$, $A_{\nu}$ is a suitable vector potential and
$J^{\mu}$ is the source of the electromagnetic field, namely charges
$\rho$ and currents $\rho{\bf v}$. In the Lorentz gauge
$A^{\mu}_{,\mu}=0$, the second pair of Maxwell's equations becomes the
wave equation with sources
\begin{equation}
{\Box}{A^{\mu}}\;=\;-4{\pi}{J^{\mu}}, \;\; {\Box}\;{\equiv}\;
{\nabla}^{2}-\frac{\partial^{2}}{\partial{t^2}}.    
\end{equation}
The equations of motion of a charge in an electromagnetic field are
\begin{equation}
m\frac{\d U^{\mu}}{\d s}\;=\;eF^{\mu\nu}U_{\nu}.  
\end{equation}
It turns out that in linearized general relativity, the field
equations in the Hilbert gauge $\psi^{\mu\nu}_{,\nu}=0$ reduce to
\begin{equation}
 \Box{\psi^{\mu\nu}}\;=\;-16\pi{\Im^{\mu\nu}}, \;\;
\psi^{\mu\nu}{\equiv}h^{\mu\nu}-\frac{1}{2}\eta_{\mu\nu}h,
\;\;h\;=\;\eta_{\mu\nu}h^{\mu\nu}  \label{wave} 
\end{equation}
and the equations of motion of the particle (for time-independent
gravitational fields) become
\begin{equation}
\frac{\d U_{\mu}}{\d s} = \left(h_{0j,\mu} -
h_{0\mu,j}\right)U^{j}+\frac{1}{2}h_{00,\mu}. 
\end{equation}
From the above, the linearized equations of general relativity
resemble those of electromagnetism with the gravito-electromagnetic
`tensor' defined as
\begin{equation}
f_{\mu\nu}\;{\equiv}\;h_{0\nu,\mu}
-h_{0\mu,\nu}.
\end{equation}
The first and second pair of the gravitational field equations (in the
Hilbert gauge) can be defined respectively as
\[
\Pi_{\mu\nu,\alpha} +\Pi_{\alpha\mu,\nu} +\Pi_{\nu\alpha,\mu}\;
=\;0,\;\;\Pi_{\mu\nu} \equiv
\left(\psi_{0\nu,\mu}-\psi_{0\mu,\nu}\right)\,,  
\]
\begin{equation}
\Box{\psi^{\mu\nu}}\;=\;-16\pi{\Im^{\mu\nu}}. \label{hgauge}
\end{equation}
It appears that it may be possible to construct a gravitational
energy-momentum (pseudo) tensor with properties similar to those of
the electromagnetic energy-momentum tensor.  However, it is worth
noting that the standard approach identifies ${\Im^{\mu\nu}}$ with the
total energy-momentum pseudo-tensor via the gauge condition
${\psi^{\mu\nu}}_{,\mu}=0$. The gravitational analog of the
electromagnetic Poynting vector is concocted in the following section.

\section{The gravitational Poynting vector}
We start by noting that the time-time and space-time components of the
second pair of the gravitational field equations $(\ref{hgauge})$ are given by
\begin{equation}
\Pi^{\mu\nu}_{,\nu}\;=\;-16\pi{\Im^{0\mu}}.  \label{gfe}
\end{equation}
The antisymmetry of $\Pi^{\mu\nu}$ yields the conserved quantity
(energy)
\begin{equation}
E\;=\;\int{\Im^{00}}\d^3x.  
\end{equation}
The Lagrangian density that generates equations $(\ref{gfe})$ is
\begin{equation}
L\;=\;-\frac{1}{64\pi}\;\Pi_{\mu\nu}\Pi^{\mu\nu}
-\rho_{m}{U_{\alpha}}B^{\alpha},
\;\;\;B^{\alpha}{\equiv}\psi^{0\alpha}. 
\label{lagrangian} 
\end{equation}
In the derivation of the field equations $(\ref{gfe})$, $B_{\alpha}$
is treated as a vector field and not as the component of a two index
object.  The energy-momentum (pseudo) tensor $\Theta^{\mu}_{\nu}$ for
the free gravitational field $(\rho_{m}=0)$ obtained upon
symmetrization of
\begin{equation}
\Theta^{\mu}_{\nu} = {\frac{\partial{B_{\sigma}}}{\partial{x^\nu}}}  
\frac{\partial{L}}{{\partial}
\left(\frac{\partial{B_{\sigma}}}{\partial{x^{\mu}}}\right)}
-\delta^{\mu}_{\nu}L,
\end{equation}
is essentially that of the free electromagnetic field
\begin{equation}
\Theta^{\mu}_{\nu} =
{\frac{1}{16\pi}}\left(\Pi^{\sigma\mu}\Pi_{\nu\sigma} 
+\frac{1}{4}\delta_{\nu}^{\mu} \;{\Pi_{\alpha\sigma}}
\;{\Pi^{\alpha\sigma}}\right). \label{poyinting} 
\end{equation}
It is shown below that the source free gravitational field equations
${\Pi^{\mu\nu}}_{,\nu}=-16\pi{\Im^{0\mu}}$ and the cyclic property
$(\ref{hgauge})$ of $\Pi_{\mu\nu,\alpha}$ give the continuity equation
${\Theta^{\mu}_{\nu}}_{,\mu}=0$ up to third order in the metric
perturbation:
\[
{\Theta^{\mu}_{\nu}}_{,\mu} =
{\frac{1}{16\pi}}\left(\Pi^{\sigma\mu}_{,\mu}\Pi_{\nu\sigma}
+\Pi^{\sigma\mu}\Pi_{\nu\sigma,\mu}+ 
\frac{1}{2} \;{\Pi_{\alpha\sigma,\nu}} \;{\Pi^{\alpha\sigma}}\right)  
\]
\[
\Pi^{\sigma\mu}\Pi_{\nu\sigma,\mu} =
{\frac{1}{2}}\Pi^{\sigma\mu}\left(\Pi_{\nu\sigma,\mu} +
\Pi_{\mu\nu,\sigma}\right) 
\]
\[
\Rightarrow \;\; 
\Pi^{\sigma\mu}\Pi_{\nu\sigma,\mu}+\frac{1}{2}
\;{\Pi_{\alpha\sigma,\nu}} \;{\Pi^{\alpha\sigma}}
= {\frac{1}{2}}\Pi^{\sigma\mu}\left(\Pi_{\nu\sigma,\mu}
+\Pi_{\mu\nu,\sigma}+\Pi_{\sigma\mu,\nu}\right) = 0 
\]
\begin{equation}
\Rightarrow \;\;{\Theta^{\mu}_{\nu}}_{,\mu} =
{\frac{1}{16\pi}}{\Pi^{\sigma\mu}_{,\mu}\Pi_{\nu\sigma}} =
{-\Im^{0\sigma}}\Pi_{\nu\sigma}. 
\label{divergence} 
\end{equation}
In vacuum $\Im^{0\sigma}$ contains only products of the derivatives of
the metric perturbation. Therefore, the right-hand-side of
$(\ref{divergence})$ is third order in $\psi^{\mu\nu}$ and can be
neglected. This yields ${\Theta^{\mu}_{\nu}}_{,\mu}=0$ to third order
in the metric perturbation. It will be shown in the following
sections that the simple Maxwell-like gravitational Poynting vector is
consistent with the radiation flux calculated with the Landau-Lifshitz
pseudo-tensor.  The gravitational Poynting vector of
$(\ref{poyinting})$ gives rise to Einstein's quadrupole radiation
formula.

\section{Linear mass quadrupole oscillator}
In this section we calculate the gravitational energy flux of a linear
mass quadrupole oscillator using the Maxwell-like gravitational
energy-momentum (pseudo) tensor $(\ref{poyinting})$. The solution for
the linear mass quadrupole oscillator is well known
\cite{Weber,Lim}. The non-vanishing components of the metric are given
by
\[
\psi^{33}\;=\;A\cos{\omega{\xi}}, \;\; {\psi^{03}} =
{\cos{\theta}}\;\psi^{33},\;\;\psi^{00} = \cos^2{\theta}\;\psi^{33}
\]
\begin{equation}
A = \frac{2\omega^{2}I}{r}, \;\; \xi\;=\;t-r.
\end{equation}
A direct calculation of the gravitational Poynting vector above leads
to
\begin{equation}
\Theta_{0}^{i} = \frac{1}{16\pi}
\left(A\omega\right)^2\sin^2{\theta}
\cos^2{\theta}\sin^{2}\left({\omega{\xi}}\right)
\;\left[\frac{x^i}{r}\right]. 
\end{equation}
The rate of change of energy averaged over one period of the
oscillation is given by
\begin{eqnarray}
\frac{\d E}{\d t}&=& -\left\langle{\oint_{\sigma}{\Theta_{0}^{i}}
\d\sigma_{i}}\right\rangle \nonumber \\ 
&=& -\frac{1}{15}I^2{\omega^6}  
\end{eqnarray}
consistent with the quadrupole radiation formula. An alternative
derivation of the gravitational quadrupole radiation formula is
provided below.

\section{The quadrupole radiation formula}
In this section we derive the quadrupole radiation formula using the
gravitational Poynting vector $(\ref{poyinting})$. The solution of the
linearized equations of general relativity $(\ref{wave})$ is given as:
\[
\psi_{ab}\;=\;-\frac{2\ddot{I}_{ab}(\xi)}{r},\;\;
\psi_{a0}\;=\;-\frac{2\ddot{I}_{ab}n_{b}}{r},
\;\;\psi_{00}\;=\;\frac{4m}{r}+\frac{2\ddot{I}_{ab}n_{a}n_{b}}{r} 
\]
\begin{equation}
n_{a}\;=\;\frac{x^a}{r},\;\; \xi\;=\;t-r,\;\;
I_{ab}(\xi)\;=\;\int{\rho}x_{a}x_{b}\d^3x \label{solution} 
\end{equation}
where $n_{a}$ is the radial unit vector and $I_{ab}$ is the moment of
inertia tensor of the system.  In $(\ref{solution})$, $\dot{A}$
denotes differentiation with respect to the retarded time $\xi$.  The
evaluation of the quadrupole radiation formula is commonly performed
with the quadrupole moment tensor
\begin{equation}
Q_{ab}\;=\;\int{\rho}{\left(3x_{a}x_{b}-{\delta}_{ab}r^2\right)}\d^3x
\label{quadrupole} 
\end{equation}
since it is symmetric and traceless. The moment of inertia tensor in
$(\ref{solution})$ relates to the quadrupole moment tensor as
\begin{equation}
I_{ab}\;=\;\frac{1}{3}Q_{ab}
+\frac{1}{3}\delta_{ab}\int{\rho{r^2}}\d^3x. \label{IQ} 
\end{equation}
The second term in $(\ref{IQ})$ doesn't contribute to radiation and
can be dropped. Therefore, the energy flux can be obtained replacing
$\psi_{\mu\nu}$ in $(\ref{solution})$ by
\begin{equation}
\Phi_{ab}\;=\;-\frac{2\ddot{Q}_{ab}}{3r},
\;\;\Phi_{a0}\;=\;-\frac{2\ddot{Q}_{ab}n_{b}}{3r},
\;\;\Phi_{00}\;=\;\frac{4m}{r}+\frac{2\ddot{Q}_{ab}n_{a}n_{b}}{3r} 
\end{equation}
The spatial derivatives of the fields $\Phi_{ab}$ involve only
derivatives of $Q_{ab}$ since the integration is carried out on a
closed surface approaching infinity. On this surface the term
proportional to $1/{r^2}$ is much smaller than that which is
proportional to $1/r$. The components of the gravito-electromagnetic
`tensor' are given below:
\[
\Pi_{\mu\nu}=\Phi_{0\nu,\mu}-\Phi_{0\mu,\nu}  
\]
\begin{equation}
\Pi_{ad}=\frac{2\stackrel{\cdots}{Q}_{db}n_{a}n_{b}}{3r} -
\frac{2\stackrel{\cdots}{Q}_{ab}n_{b}n_{d}}{3r}, \;\;  
\Pi_{0d}=\frac{2\stackrel{\cdots}{Q}_{ab}n_{a}n_{b}n_{d}}{3r}
-\frac{2\stackrel{\cdots}{Q}_{da}n_{a}}{3r}. 
\label{gravito} 
\end{equation}
The gravitational radiation flux is then given as
 \begin{equation}
16\pi\Theta_{0}^{i}\;=\;-\frac{4}{9r^2}n_{i}
\left(\stackrel{\cdots}{Q}_{db}
\stackrel{\cdots}{Q}_{kl}n_{b}n_{k}n_{l}n_{d} 
- \stackrel{\cdots}{Q}_{db} \stackrel{\cdots}{Q}_{dk}n_{b}n_{k}\right)
. \label{flux} 
\end{equation}
 The calculation of the radiated power proceeds in the usual manner by
averaging $n_{b}n_{k}n_{l}n_{d}$ and $n_{b}n_{k}$ over all spatial
directions using \cite{Landau}
\begin{equation}
\left \langle n_{b}n_{k}\right \rangle
\;=\;\frac{1}{3}\delta_{bk},\;\; \left \langle
n_{b}n_{k}n_{l}n_{d}\right \rangle
\;=\;\frac{1}{15}\left(\delta_{bk}\delta_{ld}
+\delta_{bl}\delta_{kd}+\delta_{bd}\delta_{lk}\right). 
\end{equation}
The spatial averaging of $(\ref{flux})$ yields
\begin{equation}
16\pi\Theta_{0}^{i} =
\frac{4}{45r^2}n_{i}\stackrel{\cdots}{Q}_{ab}
\stackrel{\cdots}{Q}_{ab}. 
\end{equation}
The power radiated by the system is given by the usual gravitational
quadrupole radiation formula (in S.I. units)
\begin{equation}
\frac{\d E}{\d
t}\;=\;-\oint_{\sigma}{\Theta_{0}^{i}}d\sigma_{i}
=-\frac{G}{45c^5}\left(\stackrel{\cdots}Q_{ab}\right)^2. 
\end{equation}

The symmetric energy-momentum (pseudo) tensor $(\ref{poyinting})$
allows us to define the angular momentum of the free gravitational
field in the familiar scheme
\begin{equation}
M^{\mu\nu} = \int{\left(x^{\mu}\Theta^{\nu\sigma}
-x^{\nu}\Theta^{\mu\sigma}\right)}\d S_{\sigma}  
\end{equation}
where $dS_{\sigma}$ is the element of a three dimensional space-like
hypersurface.

\section{Conclusion}
It was shown that the field equations of general relativity in the
Hilbert gauge can be derived from a Lagrangian density similar in
structure to that of the electromagnetic field. This Lagrangian
generates an energy-momentum (pseudo) tensor possessing the same
properties as the electromagnetic energy-momentum tensor. The
simple structure of the gravitational energy-momentum (pseudo) tensor 
constructed in this
manner lends itself to an elegant derivation of Einstein's quadrupole 
radiation formula. The symmetric gravitational (pseudo) tensor allows 
the definition of
angular momentum for the free gravitational field. It should be
pointed out that the formalism presented in this paper is restricted
to the Hilbert gauge. Although in principle all metrics can be transformed
into a coordinate system satisfying the Hilbert gauge condition,
future research should be aimed at constructing a gauge invariant
gravito-electromagnetic formalism.

 \section*{Acknowledgements} I would like to thank Professor W. Israel
 for his encouragement and Sean
Bohun for his careful reading of the manuscript.

\end{document}